# Empirical evaluation of construction heuristics for the multidimensional assignment problem

D. Karapetyan[*]   G. Gutin[†]   B. Goldengorin[‡]

**Abstract**

The multidimensional assignment problem (MAP) (abbreviated $s$-AP in the case of $s$ dimensions) is an extension of the well-known assignment problem. The most studied case of MAP is 3-AP, though the problems with larger values of $s$ have also a number of applications. In this paper we consider four fast construction heuristics for MAP. One of the heuristics is new. A modification of the heuristics is proposed to optimize the access to slow computer memory. The results of computational experiments for several instance families are provided and discussed.

## 1 Introduction

The Multidimensional Assignment Problem (MAP) (abbreviated $s$-AP in the case of $s$ dimensions) is a well-known optimization problem with a host of applications (see, e.g., [4, 6, 7] for 'classic' applications and [5, 17] for recent applications in solving systems of polynomial equations and centralized multisensor multitarget tracking). In fact, several applications described in [5, 6, 17] naturally require the use of $s$-AP for values of $s$ larger than 3.

MAP is an extension of a well-known Assignment Problem (AP) which is exactly two dimensional case of MAP. While AP can be solved in a polynomial time [14], $s$-AP for $s > 2$ is NP-hard [8].

For a fixed $s \geq 2$, the $s$-AP is stated as follows. Let $X_1 = X_2 = \ldots = X_s = \{1, 2, \ldots, n\}$. We will consider only vectors that belong to the Cartesian product $X = X_1 \times X_2 \times \ldots \times X_s$. Each vector $e \in X$ is assigned a non-negative weight $w(e)$. For a vector $e \in X$, the component $e_j$ denotes its $j$th coordinate, i.e., $e_j \in X_j$. A collection of $t \leq n$ vectors $e^1, e^2, \ldots, e^t$ is a *(feasible) partial assignment* if $e^i{}_j \neq e^k{}_j$ holds for each $i \neq k$ and $j \in \{1, 2, \ldots, s\}$. The

---

[*]Department of Computer Science, Royal Holloway University of London, Egham, Surrey TW20 0EX, UK, Daniel.Karapetyan@gmail.com

[†]Department of Computer Science, Royal Holloway University of London, Egham, Surrey TW20 0EX, UK, G.Gutin@rhul.ac.uk

[‡]Department of Econometrics and Operations Research, University of Groningen, P.O. Box 800, 9700 AV Groningen, The Netherlands, B.Goldengorin@rug.nl

*weight* of a partial assignment $A$ is $w(A) = \sum_{i=1}^{t} w(e^i)$. An *assignment* (or *full assignment*) is a partial assignment with $n$ vectors. The objective is to find an assignment of minimum weight.

The 3-AP is the most studied case of MAP so far. Aiex et al. introduce a Greedy Randomized Adaptive Search Procedure for 3-AP in [1]; an exact algorithm for 3-AP is proposed by Balas and Saltzman in [4]; Crama and Spieksma discuss some special cases of 3-AP and propose approximation algorithms for them. A memetic approach is tried by Huang and Lim in [11]. The more general case of $s$-AP for arbitrary values of $s$ is less studied. The most recent research by Gutin, Goldengorin and Huang overviews the previous results and discusses the worst case analysis of several MAP construction heuristics [10].

## 2 Heuristics

There are three construction heuristics for MAP known from the literature: Greedy, Max-Regret [4, 5], and ROM [10]. In this paper, we propose a modification of ROM, Shift-ROM, and compare all four heuristics with respect to solution quality and running time.

### 2.1 Greedy heuristic

The Greedy heuristic starts with an empty partial assignment $A = \emptyset$. On each of $n$ iterations Greedy finds a vector $e \in X$ of minimum weight, such that $A \cup \{e\}$ is a feasible partial assignment, and adds it to $A$.

The time complexity of Greedy heuristic is $O(n^s + (n-1)^s + \ldots + 2^s + 1) = O(n^{s+1})$ (if the Greedy algorithm is implemented via sorting of all the vectors according to their weights, the algorithm complexity is $O(n^s \cdot \log n^s)$ however this implementation is inefficient, see Subsection 3.1).

### 2.2 Max-Regret

The Max-Regret heuristic was first introduced in [4] for 3-AP and its modifications for s-AP were considered in [5].

Max-Regret proceeds as follows. Initialize partial assignment $A = \emptyset$. Set $V_d = \{1, 2, \ldots, n\}$ for each $1 \leq d \leq s$. For each dimension $d$ and each coordinate value $v \in V_d$ consider every vector $e \in X'$ such that $e_d = v$, where $X' \subset X$ is the set of 'available' vectors, i.e., $A \cup \{e\}$ is a feasible partial assignment if and only if $e \in X'$. Find two vectors $e^1{}_{\min}$ and $e^2{}_{\min}$ in the considered subset $Y_{d,v} = \{e \in X' : e_d = v\}$ such that $e^1{}_{\min} = \operatorname{argmin}_{e \in Y_{d,v}} w(e)$, and $e^2{}_{\min} = \operatorname{argmin}_{e \in Y_{d,v} \setminus \{e^1{}_{\min}\}} w(e)$. Select the pair $(d, v)$ that corresponds to the maximum difference $w(e^2{}_{\min}) - w(e^1{}_{\min})$ and add the vector $e^1{}_{\min}$ for the selected $(d, v)$ to $A$.

The time complexity of Max-Regret is $O(s \cdot n^s + s \cdot (n-1)^s + \ldots + s \cdot 2^s + s) = O(s \cdot n^{s+1})$.

## 2.3 ROM

The *Recursive Opt Matching* (ROM) is introduced in [10] as a heuristic of large domination number (see [10] for definitions and results in domination analysis). ROM proceeds as follows. Initialize the assignment $A$ with the trivial vectors: $A^i = (i, i, \ldots, i)$. On each $j$th iteration of the heuristic, $j = 1, 2, \ldots, s-1$, calculate an $n \times n$ matrix $M_{i,v} = \sum_{e \in Y(j,i,v)} w(e)$, where $Y(j, i, v)$ is a set of all vectors $e \in X$ such that the first $j$ coordinates of the vector $e$ are equal to the first $j$ coordinates of the vector $A^i$ and the $(j+1)$th coordinate of $e$ is $v$: $Y(j, i, v) = \{e \in X : e_k = A^i_k, 1 \le k \le j \text{ and } e_{j+1} = v\}$. Let permutation $\pi$ be a solution of the 2-AP for the matrix $M$. Set $A^i_{j+1} = \pi(i)$ for each $1 \le i \le n$.

The time complexity of ROM heuristic is $O((n^s + n^3) + (n^{s-1} + n^3) + \ldots + (n^2 + n^3)) = O(n^s + sn^3)$.

## 2.4 Shift-ROM

A disadvantage of the ROM heuristic is that it is not symmetric with respect to the dimensions. For example, if the vector weights do not depend significantly on the last coordinate then the algorithm is likely to work badly. Shift-ROM is intended to solve this problem by trying ROM for different permutations of the instance dimensions. However, we do not wish to try all $s!$ possible dimension permutations as that would increase the running time of the algorithm quite significantly and instead we use only $s$ different permutations: $(X_1 X_2 \ldots X_s)$, $(X_s X_1 X_2 \ldots X_{s-1})$, $(X_{s-1} X_s X_1 X_2 \ldots X_{s-2})$, $\ldots$, $(X_2 X_3 \ldots X_s X_1)$.

In other words, on each run Shift-ROM applies ROM to the problem; upon completion, it renumbers the dimensions for the next run in the following way: $X_1 := X_2$, $X_2 := X_3$, $\ldots$, $X_{s-1} := X_s$, $X_s := X_1$. After $s$ runs, the best solution is selected.

The time complexity of Shift-ROM heuristic is $O((n^s + sn^3) \cdot s) = O(sn^s + s^2 n^3)$.

## 2.5 Time Complexity Comparison

Now we can gather the information about the time complexity of the considered heuristics. The following table shows the time complexity of each of the heuristics for different values of $s$.

|  | Greedy | Max-Regret | ROM | Shift-ROM |
|---|---|---|---|---|
| Arbitrary $s$ | $O(n^{s+1})$ | $O(sn^{s+1})$ | $O(n^s + sn^3)$ | $O(sn^s + s^2 n^3)$ |
| Fixed $s = 3$ | $O(n^4)$ | $O(n^4)$ | $O(n^3)$ | $O(n^3)$ |
| Fixed $s \ge 4$ | $O(n^{s+1})$ | $O(n^{s+1})$ | $O(n^s)$ | $O(n^s)$ |

# 3 Performance Notes

Modern computer architecture is complex and, hence, not every operation takes the same time to execute. In a standard computer model it is assumed that all the operations take approximately the same time. We will use a more sophisticated model in our further discussion. The idea is to differentiate fast and low memory access operations.

The weight matrix of a MAP instance is normally stored in the Random Access Memory (RAM) of the computer. RAM's capacity is large enough for the very large instances, e.g., nowadays RAM of a common desktop PC is able to hold the weight matrix for 3-AP with $n = 700$, i.e., $3.43 \cdot 10^8$ weights[1]. RAM is a fast storage; one can load gigabytes of data from RAM in one second. However, RAM has a comparatively high latency, i.e., it takes a lot of time for the processor to access even a small portion of data in RAM. The processor cache is intended to minimize the time spent by the processor for waiting for RAM response.

The processor cache exploits two heuristics: firstly, if some data was recently used then there is a high probability it will be used again soon and, secondly, the data is used successively, i.e., if some portion of data is used now then it is probably that the successive portion of data will be used soon. As an example, consider an in place vector multiplication algorithm: on every iteration the algorithm loads a value from the memory, multiplies it and saves the result at the same memory position. So, the algorithm accesses every portion of data twice and the data is accessed successively, i.e., the algorithm accesses the first element of the data, then it accesses the second element, the third etc.

Processor cache[2] is a temporary data storage, relatively small and fast, usually located on the same chip as the processor. It contains several *cache lines* of the same size; each cache line holds a copy of some fragment of the data stored in RAM. Each time the processor needs to access some data in RAM it checks whether this data is already presented in the cache. If this is the case, it accesses this data in the cache instead. Otherwise a 'miss' is detected, the processor suspends, some cache line is freed and a new portion of data is loaded from RAM to cache. Then the processor resumes and accesses the data in the cache as normally. Note that in case of a 'miss' the system loads the whole cache line that is currently 64 bytes on the most of the modern computers [2] and this size tends to grow with the development of computer architecture. Thus, if a program accesses some value in the memory several times in a short period of time it is very probably that this data will be loaded from RAM just once and then will be stored in the cache so the access time will be minimal. Moreover, if some value is accessed and, thus, loaded from RAM to the processor cache, it is probably that the next value is also loaded since the cache line is large enough to store several values.

With respect to MAP heuristics, there are two key rules for improving the memory subsystem performance:

1. The successive access to the weight matrix (scan), i.e., access to the matrix in the order of its alignment in the memory, is strongly preferred (we use the row-major order [13] for weight matrix in our implementation of the algorithms). Note that if an algorithm accesses, e.g., every second weight in the matrix and does it in the right order, the real complexity of this scan for the memory subsystem is the same as the complexity of a full scan since loading of one value causes loading of several neighbor values.

---

[1] Here and further we assume that every weight is represented with a 4 byte integer. The calculations are provided for 2 Gb of RAM.

[2] We provide a simplified overview of cache, for detailed information, see, e.g., [3].

2. One should minimize the number of the weight matrix scans as much as possible. Even a partial matrix scan is likely to access much more data than the processor cache is able to store, i.e., the data will be loaded from RAM all over again for every scan.

Following these rules may significantly improve the running time of the heuristics. In our experiments, the benefit of following these rules was a speedup of roughly speaking 2 to 5 times.

### 3.1 Greedy heuristic optimization

A common implementation of the greedy heuristic for combinatorial optimization problem involves sorting of all the weights in the problem. In case of MAP this approach is inefficient since we actually need only $n$ vectors from the $n^s$ set. Another natural implementation of the Greedy heuristic is to scan all available vectors and to choose the lightest one on each iteration but it is very unfriendly with respect to the memory subsystem: it performs $n$ scans of the weight matrix.

We propose a combination of these approaches; our algorithm proceeds as follows. Let $A = \emptyset$ be a partial assignment and $B$ an array of vectors. While $|A| < n$, i.e., $A$ is not a full assignment, the following is repeated. We scan the weight matrix to fill array $B$ with $k$ vectors corresponding to $k$ minimal weights in non-decreasing order: if the weight of the current vector is less than the largest weight in $B$ then we insert the current vector to $B$ in an appropriate position and, if necessary, remove the last element of $B$. Then, for each vector $e \in B$, starting from the lightest, we check whether $A \cup \{e\}$ is a feasible partial assignment and, if so, add $e$ to $A$. Note, that during the second and further cycles we scan not the whole weight matrix but only a subset $X' \subset X$ of the vectors that can be included into the partial assignment $A$ with the feasibility preservation: $A \cup \{x\}$ is a partial assignment for any $x \in X'$. The size of the array $B$ is calculated as $k = \min\{64, |X'|\}$. The constant 64 is obtained from experiments.

The algorithm is especially efficient on the first iterations, i.e., in the hardest part of its work, while the most of the vectors are feasible. However, there exists a bad case for this heuristic. Assume that the weight matrix contains a lot of vectors of the minimal weight $w_{\min}$. Then the array $B$ will be filled with vectors of the weight $w_{\min}$ at the beginning of the scan and, thus, it will contain a lot of similar vectors (recall that the weight matrix is stored in the row-major order and only the last coordinates are varied at the beginning of the scan, so all the vectors processed at the beginning of the scan are likely to have the same first coordinates). As a result, selecting the first of these vectors will cause infeasibility for the other vectors in $B$. We use an additional heuristic to decrease the running time of the Greedy algorithm for such instances. Let $w_{\min}$ be the minimum possible weight: $w_{\min} = \min_{e \in X'} w(e)$ (sometimes this value is known like for Random instance family it is 1, see Section 4, or one can assume that $w_{\min} = -\infty$). If it occurs during the matrix scan that all the vectors in $B$ have the weight $w_{\min}$, i.e., $w(B_i) = w_{\min}$ for every $1 \leq i \leq k$, then the rest of the scan can be skipped because there is certainly no vector lighter than the maximum weight vector in $B$. Moreover, it is safe to update $w_{\min}$ with the maximum weight of a vector in $B$ every time before the next matrix scan.

## 3.2 Max-Regret Heuristic Optimization

The Max-Regret heuristic naturally requires $O(n^2 s)$ weight matrix partial scans. Each of these scans fixes one coordinate and, thus, every available vector $e \in X'$ (see Subsection 3.1) is accessed $s$ times during each iteration, and this access is very inefficient when the last coordinate is fixed (recall that the weight matrix is stored in a row-major order and, thus, if the last coordinate is fixed then the algorithm accesses every $n$th value in the memory, i.e., the access is very non-successive and one can assume that this scan will load the whole weight matrix from RAM to cache). In our more detailed computer model, the time complexity of the non-optimized Max-Regret is $O((s-1) \cdot n^{s+1} + n^{s+2})$.

We propose another way to implement Max-Regret. Let us scan the whole available vectors set $X'$ on each iteration. Let $L$ be an $n \times s$ matrix of the lightest vector pairs: $L^1_{i,j}$ and $L^2_{i,j}$ are the lightest vectors when the $j$th coordinate is fixed as $i$, and $w(L^1_{i,j}) \leq w(L^2_{i,j})$. To fill the matrix $L$ we do the following: for every vector $e \in X'$ and for every coordinate $1 \leq d \leq s$ check: if $w(e) < w(L^1_{e_d,d})$, set $L^2_{e_d,d} = L^1_{e_d,d}$ and $L^1_{e_d,d} = e$. Otherwise if $w(e) < L^2_{e_d,d}$, set $L^2_{e_d,d} = e$. Thus, we update the $L_{e_d,d}$ item of the matrix with the current $e$ if $w(e)$ is small enough. Having the matrix $L$, we can easily find the coordinate $d$ and the fixed value $v$ such that $w(L^2_{v,d}) - w(L^1_{v,d})$ is maximized. The vector $L^1_{v,d}$ is added to the solution and the next iteration of the algorithm is executed.

The proposed algorithm performs just $n$ partial scans of the weight matrix. The matrix $L$ is usually small enough to fit in the processor cache, so the access to $L$ is fast. Thus, the time complexity of the optimized Max-Regret in our more detailed computer model is $O(n^{s+1})$.

## 3.3 ROM Heuristic Optimization

The ROM heuristic can be implemented in a very friendly with respect to the memory access way. On the first iteration it fixes the first two coordinates ($n^2$ combinations) and enumerates all vectors with these fixed coordinates. Thus, it scans the whole weight matrix successively. On the next iteration it fixes three coordinates ($n^2$ combinations as the second coordinate depends on the first one), and enumerates all vectors with these fixed coordinates. Thus, it scans $n^2$ solid $n^{s-3}$-size fragments of the weight matrix; further iterations are similar. As a result, the time complexity of ROM in our more detailed computer model is the same as in a simple one: $O(n^s + sn^3)$.

### 3.4 Shift-ROM Heuristic Optimization

The Shift-ROM heuristic is an extension of ROM; it simply runs ROM $s$ times, starting it from different coordinates. However, not every run of ROM is efficient as a part of Shift-ROM. Let us consider the case when the first iteration of ROM fixes the last two coordinates. For each of the $n^2$ combinations of the last two coordinate values, the heuristic scans the whole weight matrix with the step $n^2$ between the accessed weights, i.e., the distance between the successively accessed weights in the memory is $n^2$ elements, which is very inefficient. A similar situation occurs when the first and the last dimensions are fixed.

To avoid this disadvantage, we propose the following algorithm. Let $M^d$ be an $n \times n$ matrix for every $1 \le d \le s$. Initialize $M^d_{i,j} = 0$ for every $1 \le d \le s$ and $1 \le i, j \le n$. For each vector $e \in X$ and for each $1 \le d \le s$ set $M^d_{e_d, e_{d+1}} = M^d_{e_d, e_{d+1}} + w(e)$ (here we assume that $e_{s+1} = e_1$). Now the matrices $M^d$ can be used for the first iteration of every ROM run.

When applying this technique, only one full matrix scan is needed for the heuristic and this scan is successive. There are several other inefficient iterations like fixing of the last three coordinates but they influence the performance insignificantly.

## 4 Test Bed

In this paper we consider four instance families.

Random instance family is a family of random instances, i.e., $w(e)$ is chosen arbitrary for each $e \in X$. Each weight is a uniformly distributed integer number in the interval $[1, 100]$. This instance family is used in [1, 4, 16] and some other papers.

Composite instance family is a family of semi-random instances. They were introduced by Crama and Spieksma for 3-AP as the $T$ problem [7]. We extend this family for $s$-AP case.

In [7] the 3-AP problem is interpreted as follows. Given a complete tripartite graph $K = \big(X_1 \cup X_2 \cup X_3, \ (X_1 \times X_2) \cup (X_1 \times X_3) \cup (X_2 \times X_3)\big)$, find a subset $A$ of $n$ triangles, $A \subset X_1 \times X_2 \times X_3$, such that every element of $X_1 \cup X_2 \cup X_3$ occurs in exactly one triangle of $A$, and the total weight of all the edges covered by triangles $A$ is minimized. The weight of a triangle is calculated as the sum of weights of its edges; the weight of an edge $(i, j) \in X_1 \times X_2$ is $d^1_{i,j}$, the weight of an edge $(i, j) \in X_2 \times X_3$ is $d^2_{i,j}$, and the weight of an edge $(i, j) \in X_1 \times X_3$ is $d^3_{i,j}$, where $d^1$, $d^2$, and $d^3$ are random $n \times n$ matrices of non-negative numbers. In our interpretation of the problem, $w(i_1, i_2, i_3) = d^1_{i_1, i_2} + d^2_{i_2, i_3} + d^3_{i_1, i_3}$.

We introduce an extension of the $T$ problem from [7]. Let us consider a graph $G\big(X_1 \cup X_2 \cup \ldots \cup X_s, \ (X_1 \times X_2) \cup (X_2 \times X_3) \cup \ldots \cup (X_{s-1} \times X_s) \cup (X_1 \times X_s)\big)$, where the weight of an edge $(i, j) \in X_1 \times X_2$ is $d^1_{i,j}$, the weight of an edge $(i, j) \in X_2 \times X_3$ is $d^2_{i,j}$, ..., and the weight of an edge $(i, j) \in X_{s-1} \times X_s$ is $d^{s-1}_{i,j}$, the weight of an edge $(i,j) \in X_1 \times X_s$ is $d_{i,j}^s$ and $d^1, d^2, \ldots, d^s$ are random $n \times n$ matrices of non-negative numbers distributed uniformly in the interval $[1, 100]$. The objective is to find a set of $n$ vertex-disjoint $s$-cycles $C \subset X_1 \times X_2 \times \ldots \times X_s$ such that the total weight of all edges covered by the cycles $C$ is minimized. In our interpretation of the problem, $w(e) = d_{e_1,e_2}^1 + d_{e_2,e_3}^2 + \ldots + d_{e_{s-1},e_s}^{s-1} + d_{e_1,e_s}^s$.

Additional conditions are applied in [7] to the random matrices $d^1$, $d^2$, and $d^3$, but we do not use these restrictions.

CS instance set is the instance set used by Crama and Spieksma in [7] for the $T\Delta$ problem that is a special case of $T$, i.e., CS is a subset of the Composite instance family. There are three types of instances, 6 instances per each type: 3 instances of size 33 and 3 instances of size 66. All the instances are of 3-AP. The CS instances meet the triangle inequality, i.e., $d^l(i,j) \leq d^l(i,k) + d^l(k,j)$ for every $l \in \{1,2,3\}$ and every $i,j,k \in \{1,2,\ldots,n\}$. For detailed information, see [7].

GP instance family contains pseudo-random instances with the predefined solutions. Predefined instances are generated by an algorithm described by Grundel and Pardalos in [9]. The generator is naturally designed for $s$-AP for arbitrary large values of $s$ and $n$. The GP generator is relatively slow and, thus, it was impossible to experiment with large GP instances.

All the instances for this paper are generated with the standard Miscrosoft .NET random generator [15] which is based on the Donald E. Knuth's subtractive random number generator algorithm [12]. For the seed of the random number sequence we use the following number: $seed = s + n + i$, where $i$ is the index of the instance of this type, $i \in \{1,2,\ldots,10\}$. The GP generator is implemented in C++ programming language and, thus, the standard Visual C++ random number generator is used instead; the seed for it is calculated in the same way. The generator for GP instances is available on the web (http://www.ici.ro/camo/forum/grudel/map.txt). The CS instances and solutions are taken from http://www.econ.kuleuven.ac.be/public/NDBAE03/instancesEJOR.htm.

## 5  Experimental Results

We have conducted a number of experiments for the optimized versions of the Greedy, Max-Regret, ROM, and Shift-ROM heuristic (see Section 3). The test bed is discussed in Section 4.

Every experiment, except the experiments with CS instances, includes 10 runs for each of the heuristics; so, 10 instances are produced for every experiment. The evaluation platform is based on AMD Athlon 64 X2 3.0 GHz processor.

The headers in the tables below are as follows:

$s$ is the number of dimensions of the instance.

$n$ is the linear size of the instance, i.e., $n = |X_1| = |X_2| = \ldots = |X_s|$.

Best is the average for the best known objective values for the corresponding instances.

Opt. is the best objective value of the instance. This header is applicable to CS instances only.

Solution error, % is the average value, in percent, over the optimal solution: $error = (value - opt)/opt \cdot 100\%$, where $value$ is the objective value obtained by the heuristic and $opt$ is the optimal objective value. For the instance families where the optimal objective value is unknown, the *Best* value (see above) is used instead.

Running time, ms is the average running time, in milliseconds.

Gr is for Greedy.

M-R is for Max-Regret.

R is for ROM.

S-R is for Shift-ROM.

The results of the experiments with the Random instance family are presented in Table 1. One can see that the solution quality of all the construction heuristics is very poor; the error exceeds 200% over the optimum value on average for every heuristic. (Note that the best values reported for the Random instances are equal or very close to the minimum possible objective values, i.e., to $n$, and, thus, are equal or very close to the optimal objective values; recall that the minimum weight of every vector in Random instance family is 1.)

Shift-ROM outperforms other heuristics with respect to solution quality on average. For some instances Max-Regret performs better but this is at the cost of much larger running times. Greedy is approximately 100 times faster than Shift-ROM and 2000 times faster than Max-Regret (one can assume that the speedup heuristic in the Greedy implementation works well in this case since Random instances have a lot of vectors of the minimum possible weight) but it is not much worse than the other heuristics with respect to solution quality.

The results of experiments with the Composite instance family are presented in Table 2. The solution quality here is much better than in the previous experiments. Max-Regret produces the best solutions for 3-AP while Shift-ROM is the best for $s$-AP for $s \geq 4$. Both Shift-ROM and especially Max-Regret are slow; the fastest heuristic is ROM and it produces relatively good solutions. Greedy is slower and produces worse solutions that the ROM heuristic for the Composite instances.

Table 3 contains the experimental results for the CS instance set. Note that CS contains instances of three types; the instances of the same type are grouped together in the table. One can see that the quality of the Shift-ROM heuristic is almost always better than the quality of all other heuristics. ROM solution quality is close to Shift-ROM solution quality and it outperforms Max-Regret and Greedy with respect to both solution quality and running time.

For the GP instance family (Table 4) Max-Regret and Shift-ROM show the best solution quality; the average error for both heuristics is about 10%. However, Shift-ROM maximum error never exceeds 16.1% while Max-Regret error reaches up to 25.7%, and Max-Regret is approximately 10 times slower than Shift-ROM. ROM is the fastest heuristic for GP and it produces only 1.5 times worse solutions than Max-Regret and Shift-ROM.

# 6 Conclusion

The comparison of the construction heuristics considered in this paper shows that the selection of a particular heuristic depends on the instance set and the quality/time requirements. Greedy is a very fast heuristic for the Random instance family; ROM and Shift-ROM perform well for the Composite instances due to their dimensionwise nature. However, in most of the cases Max-Regret and Shift-ROM, in particular, produce the best solutions. Moreover, Shift-ROM is more stable than Max-Regret with respect to both solution quality and running time. The ROM heuristic operates significantly faster than Shift-ROM at the price of relatively small solution quality decrease. The Greedy heuristic is fast but usually produces the worst solutions.

Tab. 1: Heuristics comparison for the Random instance family. Every experiment includes 10 runs.

| Inst. | Best | Solution error, % | | | | Running times, ms | | | |
|---|---|---|---|---|---|---|---|---|---|
| | | Gr | M-R | R | S-R | Gr | M-R | R | S-R |
| 3r100 | 100.0 | 101.0 | <u>18.7</u> | 67.3 | 58.2 | <u>7</u> | 814 | 9 | 46 |
| 3r150 | 150.0 | 54.5 | <u>28.8</u> | 33.6 | 30.1 | <u>14</u> | 4 260 | 26 | 147 |
| 3r200 | 200.0 | 42.0 | 16.8 | 15.5 | <u>13.8</u> | <u>25</u> | 13 070 | 64 | 367 |
| 3r250 | 250.0 | 37.4 | 21.4 | 8.4 | <u>6.4</u> | <u>36</u> | 32 043 | 128 | 714 |
| 3r300 | 300.0 | 27.2 | 13.4 | 3.8 | <u>3.3</u> | <u>39</u> | 66 719 | 184 | 1 229 |
| 3r350 | 350.0 | 25.0 | 16.8 | 2.3 | <u>1.8</u> | <u>49</u> | 113 559 | 292 | 1 889 |
| 3r400 | 400.0 | 22.2 | 7.9 | 1.5 | <u>0.9</u> | <u>55</u> | 192 683 | 417 | 2 827 |
| 3r450 | 450.0 | 20.9 | 8.7 | 0.6 | <u>0.3</u> | <u>68</u> | 309 674 | 635 | 4 115 |
| 4r20 | 20.8 | 208.7 | <u>185.6</u> | 310.6 | 261.5 | <u>1</u> | 29 | 1 | 11 |
| 4r30 | 30.0 | 206.3 | <u>193.3</u> | 229.7 | 203.3 | <u>3</u> | 204 | 6 | 48 |
| 4r40 | 40.0 | 188.8 | <u>118.8</u> | 199.8 | 158.5 | <u>4</u> | 825 | 18 | 140 |
| 4r50 | 50.0 | 105.2 | <u>93.0</u> | 128.2 | 118.4 | <u>7</u> | 2 650 | 51 | 349 |
| 4r60 | 60.0 | 107.7 | <u>98.0</u> | 115.2 | 100.0 | <u>8</u> | 6 355 | 94 | 713 |
| 4r70 | 70.0 | 85.6 | <u>68.7</u> | 91.4 | 85.6 | <u>9</u> | 13 176 | 151 | 1 357 |
| 4r80 | 80.0 | 74.0 | <u>48.8</u> | 76.4 | 71.1 | <u>12</u> | 27 285 | 278 | 2 214 |
| 4r90 | 90.0 | <u>38.3</u> | 60.1 | 70.2 | 57.7 | <u>12</u> | 45 485 | 390 | 3 746 |
| 5r10 | 10.5 | 545.7 | 459.0 | 641.9 | <u>363.8</u> | <u>1</u> | 12 | 1 | 8 |
| 5r15 | 15.0 | 359.3 | 338.0 | 405.3 | <u>333.3</u> | <u>2</u> | 130 | 6 | 59 |
| 5r20 | 20.0 | 232.0 | <u>221.5</u> | 332.0 | 235.5 | <u>2</u> | 663 | 23 | 237 |
| 5r25 | 25.0 | 229.2 | <u>156.8</u> | 249.2 | 219.2 | <u>3</u> | 2 448 | 69 | 719 |
| 5r30 | 30.0 | 228.7 | 220.7 | 230.0 | <u>185.3</u> | <u>4</u> | 7 057 | 161 | 1 718 |
| 5r35 | 35.0 | 155.1 | <u>137.1</u> | 194.6 | 165.7 | <u>5</u> | 16 552 | 337 | 3 707 |
| 6r6 | 6.5 | 838.5 | 923.1 | 912.3 | <u>452.3</u> | <u>0</u> | 5 | 0 | 5 |
| 6r9 | 9.0 | 674.4 | 548.9 | 558.9 | <u>387.8</u> | <u>1</u> | 63 | 4 | 49 |
| 6r12 | 12.0 | 372.5 | 392.5 | 440.8 | <u>315.8</u> | <u>2</u> | 417 | 22 | 292 |
| 6r15 | 15.0 | 336.7 | 388.0 | 408.0 | <u>313.3</u> | <u>2</u> | 1 599 | 83 | 1 037 |
| 6r18 | 18.0 | 321.1 | <u>271.1</u> | 372.8 | 282.8 | <u>3</u> | 5 778 | 211 | 3 003 |
| 7r4 | 4.3 | 1134.9 | 1337.2 | 651.2 | <u>393.0</u> | 0 | 2 | <u>0</u> | 2 |
| 7r6 | 6.0 | 788.3 | 768.3 | 851.7 | <u>446.7</u> | <u>1</u> | 28 | 2 | 33 |
| 7r8 | 8.0 | 726.3 | 761.3 | 626.3 | <u>403.8</u> | <u>1</u> | 218 | 16 | 239 |
| 7r10 | 10.0 | 722.0 | 663.0 | 541.0 | <u>355.0</u> | <u>1</u> | 1 176 | 78 | 986 |
| 7r12 | 12.0 | 420.8 | 479.2 | 475.0 | <u>345.8</u> | <u>2</u> | 4 788 | 261 | 3 757 |
| 8r4 | 4.0 | 927.5 | 1185.0 | 1162.5 | <u>510.0</u> | <u>0</u> | 6 | 1 | 9 |
| 8r6 | 6.0 | 615.0 | 688.3 | 790.0 | <u>473.3</u> | <u>1</u> | 184 | 14 | 218 |
| 8r8 | 8.0 | 548.8 | 477.5 | 723.8 | <u>433.8</u> | <u>1</u> | 1 987 | 145 | 2 097 |
| All avg. | | 329.2 | 326.1 | 340.6 | <u>222.5</u> | <u>11</u> | 24 913 | 119 | 1 088 |
| 3-AP avg. | | 41.3 | 16.6 | 16.6 | <u>14.3</u> | <u>37</u> | 91 603 | 219 | 1 417 |
| 4-AP avg. | | 126.8 | <u>108.3</u> | 152.7 | 132.0 | <u>7</u> | 12 001 | 124 | 1 072 |
| 5-AP avg. | | 291.7 | 255.5 | 342.2 | <u>250.5</u> | <u>3</u> | 4 477 | 99 | 1 075 |
| 6-AP avg. | | 508.6 | 504.7 | 538.6 | <u>350.4</u> | <u>2</u> | 1 572 | 64 | 877 |
| 7-AP avg. | | 758.5 | 801.8 | 629.0 | <u>388.9</u> | <u>1</u> | 1 242 | 71 | 1 003 |
| 8-AP avg. | | 697.1 | 783.6 | 892.1 | <u>472.4</u> | <u>1</u> | 726 | 53 | 775 |

Tab. 2: Heuristics comparison for the Composite instance family. Every experiment includes 10 runs.

| Inst. | Best | Solution error, % | | | | Running times, ms | | | |
|---|---|---|---|---|---|---|---|---|---|
| | | Gr | M-R | R | S-R | Gr | M-R | R | S-R |
| 3c100 | 1396.8 | 43.4 | <u>27.2</u> | 40.0 | 33.8 | 15 | 870 | <u>9</u> | 60 |
| 3c150 | 1760.2 | 38.8 | <u>22.3</u> | 33.4 | 30.1 | 62 | 4 391 | <u>27</u> | 162 |
| 3c200 | 2017.8 | 37.2 | <u>19.1</u> | 35.6 | 33.5 | 158 | 12 979 | <u>68</u> | 401 |
| 3c250 | 2276.1 | 30.4 | <u>17.0</u> | 35.1 | 33.6 | 292 | 31 738 | <u>129</u> | 749 |
| 3c300 | 2551.4 | 27.5 | <u>13.2</u> | 34.7 | 32.6 | 557 | 69 101 | <u>211</u> | 1 229 |
| 3c350 | 2696.4 | 30.4 | <u>13.1</u> | 37.9 | 36.8 | 916 | 125 919 | <u>314</u> | 2 023 |
| 3c400 | 3008.5 | 27.0 | <u>9.7</u> | 34.6 | 33.2 | 1 424 | 211 784 | <u>480</u> | 2 925 |
| 3c450 | 3222.1 | 24.6 | <u>9.2</u> | 35.1 | 33.4 | 2 017 | 338 248 | <u>685</u> | 4 282 |
| 4c20 | 875.7 | 40.8 | 32.7 | 24.8 | <u>16.7</u> | 2 | 34 | <u>1</u> | 10 |
| 4c30 | 930.1 | 51.3 | 38.0 | 27.0 | <u>21.2</u> | 10 | 222 | <u>6</u> | 48 |
| 4c40 | 1040.0 | 50.3 | 41.2 | 32.3 | <u>27.2</u> | 30 | 919 | <u>19</u> | 145 |
| 4c50 | 1139.1 | 58.7 | 40.4 | 38.1 | <u>30.3</u> | 83 | 2 700 | <u>47</u> | 356 |
| 4c60 | 1251.0 | 53.0 | 35.6 | 32.8 | <u>27.0</u> | 154 | 6 760 | <u>93</u> | 721 |
| 4c70 | 1360.7 | 48.5 | 33.9 | 31.5 | <u>27.0</u> | 287 | 14 678 | <u>173</u> | 1 332 |
| 4c80 | 1449.5 | 47.8 | 33.7 | 30.8 | <u>27.5</u> | 543 | 28 037 | <u>284</u> | 2 259 |
| 4c90 | 1544.9 | 45.2 | 25.4 | 29.3 | <u>25.3</u> | 825 | 51 083 | <u>457</u> | 3 672 |
| 5c10 | 812.6 | 38.2 | 26.0 | 12.1 | <u>7.9</u> | 2 | 14 | <u>1</u> | 8 |
| 5c15 | 923.1 | 41.7 | 33.4 | 19.8 | <u>10.6</u> | 9 | 147 | <u>7</u> | 57 |
| 5c20 | 988.8 | 53.2 | 40.2 | 21.2 | <u>16.3</u> | 36 | 697 | <u>24</u> | 226 |
| 5c25 | 1026.0 | 54.2 | 46.9 | 26.6 | <u>19.5</u> | 117 | 2 577 | <u>71</u> | 699 |
| 5c30 | 1091.5 | 63.0 | 47.5 | 30.2 | <u>23.7</u> | 276 | 7 168 | <u>168</u> | 1 666 |
| 5c35 | 1171.9 | 60.5 | 47.8 | 28.1 | <u>20.8</u> | 583 | 18 804 | <u>381</u> | 3 680 |
| 6c6 | 817.6 | 23.5 | 21.0 | 9.0 | <u>3.5</u> | 1 | 5 | <u>1</u> | 5 |
| 6c9 | 911.4 | 33.2 | 28.8 | 11.1 | <u>6.1</u> | 7 | 66 | <u>4</u> | 52 |
| 6c12 | 1025.9 | 38.4 | 31.4 | 15.1 | <u>9.9</u> | 39 | 412 | <u>23</u> | 273 |
| 6c15 | 1011.1 | 44.4 | 39.4 | 19.7 | <u>14.3</u> | 136 | 1 956 | <u>99</u> | 989 |
| 6c18 | 1073.7 | 49.5 | 44.1 | 18.5 | <u>14.8</u> | 387 | 6 657 | <u>246</u> | 2 919 |
| 7c4 | 757.4 | 16.5 | 12.0 | 5.2 | <u>1.2</u> | 0 | 2 | <u>0</u> | 2 |
| 7c6 | 940.9 | 29.0 | 23.4 | 5.6 | <u>2.3</u> | 4 | 29 | <u>2</u> | 32 |
| 7c8 | 1000.4 | 29.3 | 26.7 | 12.9 | <u>5.1</u> | 25 | 248 | <u>17</u> | 240 |
| 7c10 | 1086.1 | 36.8 | 34.2 | 11.0 | <u>5.6</u> | 124 | 1 278 | <u>92</u> | 998 |
| 7c12 | 1159.3 | 45.9 | 39.5 | 15.2 | <u>8.4</u> | 438 | 5 048 | <u>289</u> | 3 703 |
| 8c4 | 841.3 | 15.9 | 12.5 | 4.0 | <u>1.0</u> | 1 | 6 | <u>1</u> | 9 |
| 8c6 | 1074.1 | 26.8 | 24.0 | 4.5 | <u>1.8</u> | 23 | 183 | <u>14</u> | 218 |
| 8c8 | 1148.3 | 34.2 | 31.7 | 10.6 | <u>4.5</u> | 203 | 2 039 | <u>129</u> | 2 069 |
| All avg. | | 39.7 | 29.2 | 23.2 | <u>18.5</u> | 280 | 27 051 | <u>131</u> | 1 092 |
| 3-AP avg. | | 32.4 | <u>16.4</u> | 35.8 | 33.4 | 680 | 99 379 | <u>240</u> | 1 479 |
| 4-AP avg. | | 49.4 | 35.1 | 30.8 | <u>25.3</u> | 242 | 13 054 | <u>135</u> | 1 068 |
| 5-AP avg. | | 51.8 | 40.3 | 23.0 | <u>16.5</u> | 171 | 4 901 | <u>109</u> | 1 056 |
| 6-AP avg. | | 37.8 | 32.9 | 14.7 | <u>9.7</u> | 114 | 1 819 | <u>75</u> | 848 |
| 7-AP avg. | | 31.5 | 27.2 | 10.0 | <u>4.5</u> | 118 | 1 321 | <u>80</u> | 995 |
| 8-AP avg. | | 25.6 | 22.7 | 6.4 | <u>2.4</u> | 76 | 743 | <u>48</u> | 765 |

Tab. 3: Heuristics comparison for the CS instance family.

| Inst. | $n$ | Opt. | Solution error, % | | | | Running times, ms | | | |
|---|---|---|---|---|---|---|---|---|---|---|
| | | | Gr | M-R | R | S-R | Gr | M-R | R | S-R |
| 3DA99N1 | 33 | 1608.0 | 24.5 | 19.9 | <u>0.6</u> | <u>0.6</u> | 0.9 | 11.4 | <u>0.4</u> | 1.8 |
| 3DA99N2 | 33 | 1401.0 | 19.3 | 10.3 | 1.1 | <u>0.8</u> | 0.8 | 11.5 | <u>0.4</u> | 2.0 |
| 3DA99N3 | 33 | 1604.0 | 15.3 | 15.3 | <u>0.3</u> | <u>0.3</u> | 0.7 | 11.0 | <u>0.4</u> | 1.9 |
| 3DA198N1 | 66 | 2662.0 | 23.7 | 17.0 | 1.1 | <u>0.2</u> | 4.9 | 156.0 | <u>2.8</u> | 14.5 |
| 3DA198N2 | 66 | 2449.0 | 33.1 | 36.0 | 2.0 | <u>0.9</u> | 5.7 | 187.2 | <u>2.8</u> | 13.8 |
| 3DA198N3 | 66 | 2758.0 | 17.4 | 26.0 | 1.6 | <u>0.6</u> | 4.9 | 156.0 | <u>2.7</u> | 17.6 |
| 3DIJ99N1 | 33 | 4797.0 | 6.6 | 4.8 | 1.8 | <u>1.4</u> | 2.4 | 11.3 | <u>0.4</u> | 2.2 |
| 3DIJ99N2 | 33 | 5067.0 | 5.6 | 3.3 | 1.9 | <u>1.3</u> | 2.2 | 11.2 | <u>0.5</u> | 2.4 |
| 3DIJ99N3 | 33 | 4287.0 | 7.0 | 6.2 | <u>1.3</u> | <u>1.3</u> | 1.8 | 11.3 | <u>0.4</u> | 2.1 |
| 3DI198N1 | 66 | 9684.0 | 6.1 | 4.4 | 1.4 | <u>0.9</u> | 15.7 | 124.8 | <u>2.8</u> | 15.6 |
| 3DI198N2 | 66 | 8944.0 | 6.9 | 4.9 | 2.1 | <u>2.1</u> | 17.7 | 140.4 | <u>3.0</u> | 17.2 |
| 3DI198N3 | 66 | 9745.0 | 7.0 | 6.2 | 1.8 | <u>1.2</u> | 16.7 | 171.6 | <u>3.0</u> | 15.3 |
| 3D1299N1 | 33 | 133.0 | 6.8 | 5.3 | 4.5 | <u>1.5</u> | 0.6 | 10.5 | <u>0.3</u> | 1.7 |
| 3D1299N2 | 33 | 131.0 | 8.4 | 3.8 | 6.1 | <u>3.1</u> | 0.6 | 10.3 | <u>0.4</u> | 1.8 |
| 3D1299N3 | 33 | 131.0 | 7.6 | <u>3.1</u> | <u>3.1</u> | <u>3.1</u> | 0.5 | 10.3 | <u>0.3</u> | 1.7 |
| 3D1198N1 | 66 | 286.0 | 5.9 | <u>2.8</u> | 3.1 | 3.1 | 4.6 | 156.0 | <u>2.5</u> | 13.2 |
| 3D1198N2 | 66 | 286.0 | 3.1 | 3.5 | 3.1 | <u>2.4</u> | 3.9 | 156.0 | <u>2.6</u> | 13.5 |
| 3D1198N3 | 66 | 282.0 | 7.4 | <u>2.5</u> | 4.3 | 3.9 | 4.3 | 140.4 | <u>2.3</u> | 13.1 |
| All avg. | | | 11.8 | 9.7 | 2.3 | <u>1.6</u> | 4.9 | 82.6 | <u>1.6</u> | 8.4 |

Tab. 4: Heuristics comparison for the GP instance family. Every experiment includes 10 runs.

|  |  | Solution error, % | | | | Running times, ms | | | |
|---|---:|---:|---:|---:|---:|---:|---:|---:|---:|
| Inst. | Opt. | Gr | M-R | R | S-R | Gr | M-R | R | S-R |
| 3gp20 | 98.8 | 17.2 | 18.1 | 18.0 | <u>16.1</u> | 0.4 | 1.7 | <u>0.1</u> | 0.4 |
| 3gp30 | 150.9 | <u>11.0</u> | 11.9 | 14.1 | 13.8 | 1.0 | 7.7 | <u>0.3</u> | 1.4 |
| 3gp40 | 197.4 | 12.4 | <u>10.5</u> | 13.9 | 13.6 | 2.0 | 23.4 | <u>0.7</u> | 3.1 |
| 3gp50 | 251.5 | 9.9 | <u>9.6</u> | 12.2 | 11.3 | 4.0 | 59.0 | <u>1.6</u> | 6.3 |
| 3gp60 | 286.1 | 8.9 | <u>8.4</u> | 13.1 | 12.1 | 6.4 | 126.7 | <u>1.9</u> | 11.9 |
| 3gp70 | 343.0 | 8.3 | <u>7.7</u> | 12.1 | 11.4 | 10.6 | 215.3 | <u>3.1</u> | 16.9 |
| 3gp80 | 403.7 | 7.6 | <u>6.7</u> | 11.6 | 10.7 | 17.5 | 368.2 | <u>4.7</u> | 25.3 |
| 3gp90 | 434.5 | 7.5 | <u>5.9</u> | 11.4 | 10.5 | 25.9 | 578.8 | <u>6.4</u> | 35.8 |
| 3gp100 | 504.4 | <u>5.6</u> | 5.8 | 9.8 | 9.4 | 40.0 | 803.4 | <u>8.7</u> | 47.4 |
| 4gp10 | 51.5 | 22.5 | 19.8 | 17.7 | <u>12.6</u> | 0.5 | 1.4 | <u>0.1</u> | 0.8 |
| 4gp15 | 69.6 | 16.8 | 13.9 | 9.2 | <u>7.0</u> | 1.6 | 7.9 | <u>0.5</u> | 3.3 |
| 4gp20 | 106.1 | 11.7 | 10.9 | 5.3 | <u>4.5</u> | 5.5 | 31.5 | <u>1.4</u> | 10.0 |
| 4gp25 | 132.9 | 8.1 | 8.3 | 3.5 | <u>3.1</u> | 15.4 | 94.5 | <u>3.2</u> | 27.6 |
| 4gp30 | 145.2 | 8.8 | 8.9 | 2.2 | <u>1.5</u> | 34.5 | 205.9 | <u>6.7</u> | 50.6 |
| 5gp4 | 20.1 | 30.3 | 23.4 | 31.8 | <u>15.9</u> | 0.2 | 0.3 | <u>0.0</u> | 0.3 |
| 5gp6 | 26.9 | 32.0 | <u>13.8</u> | 21.2 | 14.5 | 0.4 | 0.8 | <u>0.2</u> | 0.8 |
| 5gp8 | 36.3 | 28.7 | 23.1 | 17.6 | <u>12.7</u> | 0.9 | 3.7 | <u>0.3</u> | 2.8 |
| 5gp10 | 49.6 | 17.7 | 10.1 | 9.5 | <u>7.1</u> | 2.7 | 13.2 | <u>0.8</u> | 8.5 |
| 5gp12 | 66.2 | 12.8 | 8.9 | 9.2 | <u>6.3</u> | 6.4 | 36.1 | <u>2.0</u> | 19.1 |
| 6gp4 | 20.4 | 23.5 | <u>0.0</u> | 28.4 | 14.2 | 0.3 | 0.4 | <u>0.1</u> | 0.5 |
| 6gp6 | 30.3 | 30.4 | <u>8.9</u> | 18.8 | 11.9 | 1.5 | 5.4 | <u>0.6</u> | 4.5 |
| 6gp8 | 41.8 | 24.6 | <u>1.4</u> | 13.6 | 8.6 | 5.2 | 33.0 | <u>2.3</u> | 25.8 |
| 7gp2 | 10.9 | 47.7 | 25.7 | 13.8 | <u>5.5</u> | 0.1 | 0.1 | <u>0.0</u> | 0.2 |
| 7gp3 | 13.8 | 33.3 | <u>10.1</u> | 34.8 | 13.8 | 0.2 | 0.3 | <u>0.0</u> | 0.5 |
| 7gp4 | 20.2 | 31.7 | <u>0.0</u> | 30.2 | 13.4 | 0.4 | 1.6 | <u>0.2</u> | 2.2 |
| 7gp5 | 25.6 | 27.0 | <u>5.9</u> | 19.5 | 8.6 | 1.4 | 8.5 | <u>0.8</u> | 8.7 |
| 8gp2 | 9.9 | 36.4 | 10.1 | 19.2 | <u>7.1</u> | 0.1 | 0.1 | <u>0.0</u> | 0.3 |
| 8gp3 | 16.3 | 54.0 | <u>0.0</u> | 23.9 | 11.7 | 0.3 | 0.7 | <u>0.2</u> | 1.3 |
| 8gp4 | 19.2 | 21.4 | <u>6.8</u> | 28.1 | 8.3 | 1.0 | 8.2 | <u>0.8</u> | 8.8 |
| All avg. |  | 21.0 | <u>10.2</u> | 16.3 | 10.2 | 6.4 | 91.0 | <u>1.6</u> | 11.2 |
| 3-AP avg. |  | 9.8 | <u>9.4</u> | 12.9 | 12.1 | 12.0 | 242.7 | <u>3.0</u> | 16.5 |
| 4-AP avg. |  | 13.6 | 12.4 | 7.6 | <u>5.8</u> | 11.5 | 68.2 | <u>2.4</u> | 18.5 |
| 5-AP avg. |  | 24.3 | 15.9 | 17.9 | <u>11.3</u> | 2.1 | 10.8 | <u>0.7</u> | 6.3 |
| 6-AP avg. |  | 26.2 | <u>3.4</u> | 20.3 | 11.6 | 2.3 | 13.0 | <u>1.0</u> | 10.3 |
| 7-AP avg. |  | 34.9 | 10.4 | 24.6 | <u>10.3</u> | 0.5 | 2.6 | <u>0.3</u> | 2.9 |
| 8-AP avg. |  | 37.2 | <u>5.6</u> | 23.7 | 9.0 | 0.5 | 3.0 | <u>0.3</u> | 3.5 |